\documentclass[prl,floats,twocolumn,aps,superscriptaddress,showpacs]{revtex4}
\usepackage{amsmath}
\usepackage{graphicx}
\usepackage{epsfig}

\newcommand{\vp}{{\bf  v}}
\newcommand{\vf}{{\bf  u}}
\newcommand{\bx}{{\bf  x}}

\newcommand{\beqn}{\begin{eqnarray}}
\newcommand{\eeqn}{\end{eqnarray}}

\begin{document}

\title{Sand stirred by chaotic advection}

\author{Crist\'obal L\'opez}
\affiliation{
 Dipartimento di Fisica,
Universit\`a di Roma `La Sapienza',  Piazzale A. Moro 2, I-00185, Rome,
Italy. 
}
\author{Andrea Puglisi}
\affiliation{
 Dipartimento di Fisica,
Universit\`a di Roma `La Sapienza',  Piazzale A. Moro 2, I-00185, Rome,
Italy. 
}
\affiliation{INFM Center for Statistical Mechanics and Complexity, Italy.} 
\date{\today}

\begin{abstract}
We study the spatial structure of a granular material,
$N$ particles subject to inelastic mutual collisions, when it is
stirred by a bidimensional smooth chaotic flow. A simple dynamical
model is introduced where four different time scales are explicitly
considered: i) the Stokes time, accounting for the inertia of the
particles, ii) the mean collision time among the grains, iii) the
typical time scale of the flow, and iv) the inverse of the Lyapunov
exponent of the chaotic flow, which gives a typical time for the
separation of two initially close parcels of fluid.  Depending on the
relative values of these different times a complex scenario appears
for the long-time steady spatial distribution of particles, where
clusters of particles may or not appear.
\end{abstract}
\maketitle



Many different physical processes can be studied in the general
framework of {\it transport of finite-size particles by an external
flow}.  Sand suspended on the surface of a river or driven by the
wind, chemical pollutants advected by atmospheric flows, and
impurities driven by oceanic currents, are just but a few of the many
important real examples that fit into this.

Recently, much effort coming from the {\it Chaotic
Advection}~\cite{gen:adv} and {\it Turbulence}~\cite{gen:tur}
communities has been devoted to the study of advection of non-reacting
inertial particles by chaotic or turbulent flows. Though some works
considered the effect of {\it elastic collisions}, see for
example~\cite{collins}, the obvious fact that in real systems
collisions among particles dissipate kinetic energy into heat ({\it
inelastic collisions}) has not been, to our knowledge, much subject of
study. From another perspective, the so-called {\it Granular Media}
community has also widely considered~\cite{granular} the properties of
particles colliding inelastically, and externally driven, for example,
by a shear flow or by vibrating periodically the container of the
particles
\cite{campbell}.  However, the influence of an external chaotic or
turbulent flow seems to be overcome in the literature (see, however,
\cite{shinbrot1} for the case of densely packed granular systems).

In this paper, we try to investigate numerically the full problem.
Equivalently, we study the effect of collisions on inertial particles
immersed in a flow showing chaotic advection~\cite{aref}, or rather,
the properties of a dilute granular system (in the literature
the term {\em granular gas}~\cite{gas} is widely used) under chaotic
stirring. It is by now clear that transport behavior in the so-called
Batcherlor's regime, that is, in the range of scales between the
smallest typical length scale of the velocity field and the
characteristic diffusion length scale, is equivalent in a turbulent
flow and a chaotic advection flow. 

We introduce a simple model of granular material consisting of many
particles subject to inelastic mutual collisions, and immersed in a
smooth two-dimensional chaotic flow without gravity.  Granular
particles are assumed to be much heavier than fluid. In particular, we
focus on the steady spatial structure of the system. It is known, in
fact, that the presence of inertia can induce preferential
concentration~\cite{gen:adv,gen:tur}. One may ask what happens if the
diffusive role of collisions (in the elastic limit) among particles is
taken into account. At the same time is also well known that
dissipative collisions originate particle
clustering~\cite{puglio}. Therefore, in spite of its simplicity, we
will see that the model shows many transitions from clustering to
unclustering depending on the interplay between the inertia of the
particles, the chaotic flow and the collisions.

We consider $N$ identical particles of mass $m=1$ on a two-dimensional
domain $L \times L$ (with periodic boundary conditions) driven by an external
velocity field $\vf (\bx,t)$

\begin{subequations}
\label{modelo}
\begin{align}
\frac{d\vp_i (t)}{dt}&=-\frac{1}{\tau}\left( \vp_i (t)-\vf (\bx_i(t),t)\right),
\label{modelo1} \\
\frac{d\bx_i (t)}{dt}&=\vp_i (t),
\label{modelo2}
\end{align}
\end{subequations}
where $i=1,...,N$, $\vp_i$ is the velocity of the particle $i$, $\tau$
is the standard Stokes time, that depends on the particles diameter,
viscosity, and densities of fluid and particles.  Hence the term in
the r.h.s of Eq.~(\ref{modelo1}) is simply a viscous Stokes drag. In
addition the particles are subject to binary instantaneous inelastic
collisions according to the rule

\begin{equation}
\label{inelastic_collision}
\mathbf{v}_{i(j)}' =\mathbf{v}_{i(j)}-\frac{1+r}{2}((\mathbf{v}_{i(j)}-\mathbf{v}_{j(i)})
\cdot \hat{\mathbf{n}}) \hat{\mathbf{n}}
\end{equation}
where $i$ and $j$ are the indexes of colliding particles, the primes
denote the velocities after collision and $r \in [0,1]$ is the
restitution coefficient ($r=1$ for the elastic case). This rule ensures
the conservation of momentum, while the kinetic energy is dissipated
if $r<1$. 

Some further clarifications on the model are worth mentioning. In the
absence of collisions the equations in \eqref{modelo} are simply
the equations of motion of a rigid sphere in a flow where the Faxen
corrections, the added mass term, and the Bernoulli term are
neglected~\cite{maxey}.  This is a consistent approximation when the
particles of granular material are much heavier than that of the
fluid, which is the case we are considering. Moreover,
we adopt the Direct Simulation Monte Carlo (DSMC) scheme to integrate
the dynamics of the system~\cite{bird}. The DSMC algorithm, widely used
in numerical hydrodynamics, consists of a time discretization with
fixed time step: in every time step there is a free flow-step and a
collision step.  In the free-flow step the motion of particles is
integrated according to equations~\eqref{modelo}
without taking into account the possible collisions. In the collision
step every particle has a probability of colliding with neighboring
particles (in a disk with a radius smaller than the mean free path):
the probability of particle $i$ to collide with particle $j$ is
proportional to the relative velocity
$|\mathbf{v}_i-\mathbf{v}_j|$. This scheme has been proved to converge
to the solution of the Boltzmann equation of the corresponding hard
disk gas and its key feature is (through the randomization of
collisions) the assumption of Molecular Chaos, i.e. lack of
correlations for colliding particles,
$P(\mathbf{v}_i,\mathbf{v}_j,t)=P(\mathbf{v}_i,t)P(\mathbf{v}_j,t)$. It
is important to note that this assumption does not rule out
correlations at scales larger than the scale of a diameter, and the
DSMC is often used to study vortices~\cite{vortices} and other
instabilities. In the context of inelastic gases it has been
previously used also for the investigation of the clustering
phenomenon~\cite{puglio}.

A statistically stationary state is reached when the dissipation of
energy because of the inelastic collisions is balanced with the
continuous energy injection coming from the external
flow~\cite{note}. We determine that the statistically stationary state
is reached when the typical fluctuations of the total energy of the
system, $E=\langle v_i^2 \rangle$ (the so-called {\em granular
temperature}), are rather small, typically some small percentage of
$E$. As already mentioned, the properties of this state are the
central objective of this work, with special emphasis to the
clusterization of particles.

The formal introduction of the collision operator and
adimensionalization of the dynamical system
Eqs.~(\ref{modelo1},\ref{modelo2}) will help in order to see the relevance of
the distinct terms.  If we denote $T$ as the typical time of the flow,
$L$ the typical length and $\bf F_{col}/\tau_c$ the collision
operator, with $\tau_c$ the mean collision time, we can rewrite
Eq.~(\ref{modelo1})  as follows
\begin{equation}
\frac{\tau}{T}\frac{d\hat\vp_i}{d\hat t}=-(\hat \vp_i-\hat \vf)+
\frac{\tau}{\tau_c}\bf F_{col},
\label{adim}
\end{equation}
where $\hat t=t/T$, $\hat \vp_i=\vp_i T/L$, and $\hat \vf =\vf T/L$.

For our numerical studies we use the time periodic flow 
$\vf (\bx,t) =(u_x (x,y,t), u_y (x,y,t))$
 given by:
\beqn
u_x &=& 2\pi L \alpha /T \sin(2 \pi y/L), \nonumber \\
u_y&=&0,
\eeqn
if $t$ mod $T < T/2$, and
\beqn
u_x&=&0, \nonumber \\
u_y&=&-2 \pi L \alpha /T \sin(2 \pi x/L),
\eeqn
if $t$ mod $T >T/2$.

Periodic boundary conditions are assumed, and for $\alpha >0.4$ a
unique chaotic region without KAM tori (numerically distinguishable)
is obtained~\cite{berthier}.  The Lyapunov exponent for the flow is
given by $\lambda = 1.96 \ln (3.35 \alpha)/T$ and introduces a new
time scale, $\tau_f=1/\lambda$, for the exponential separation rate of
close fluid parcels. Our results has been checked to be independent of
the kind of flow. In particular, we have alternatively used the
cellular flow derived from the stream function
$\Psi(x,y,t)=U\sin(\frac{n \pi x + B \cos (\omega t)}{L})\sin(\frac{n
\pi y}{L})$, where $U$ is the maximal velocity of the flow, $n$ the
number of cells, $B$ the strength of the time dependent forcing and
$\omega$ its frequency~\cite{gollub}.

Depending on the relative values of the above introduced time scales:
$\tau$, $T$, $\tau_f$ and $\tau_c$, a very complex scenario appears
for the spatial distribution of the particles in the stationary state.
Some preliminary general results can be anticipated.  For example, it
seems obvious that when $\tau_f <\tau_c$ the exponential separation
produced by the chaotic driving of the fluid may avoid any
clusterization. The opposite may happen if $\tau_c < \tau_f$ as the
mechanism of aggregation due to inelastic collisions resists the
dispersing effect of the chaotic flow. However, as we will see it is
also very important to take into account the compaction due to the
inertia.  In any case, discussing all the different possibilities
(taking also into account some values of the parameter $r$) goes
beyond the scope of this Paper. Thus, we just restrict ourselves to
study some of the most interesting cases.

\begin{figure}[htb]
\begin{center}
\includegraphics[clip=true,width=\columnwidth, keepaspectratio]{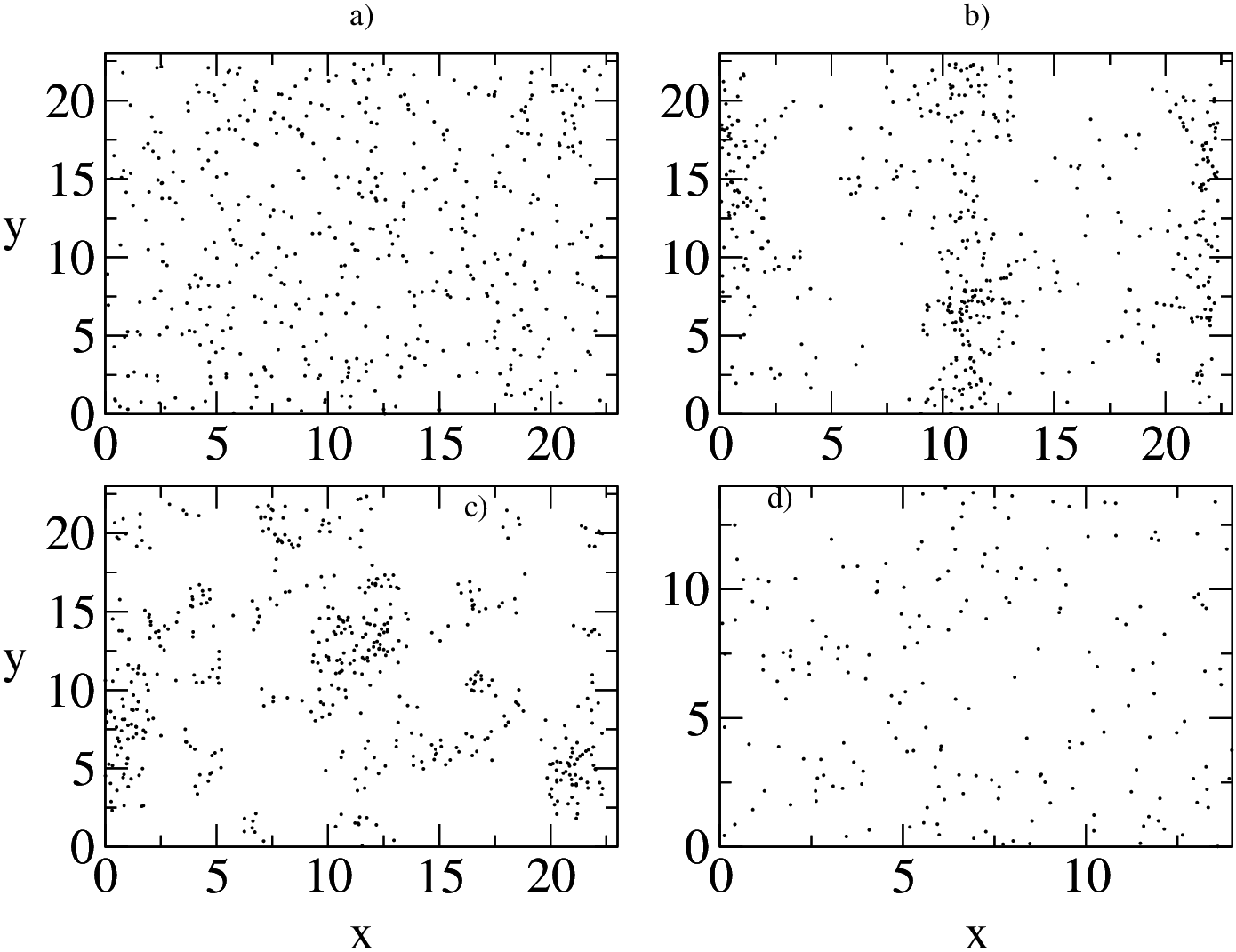}
\end{center}
\caption{Spatial distribution of particles in the steady state, when
the inertia is irrelevant (or almost), $\tau <T$.
a) Case with no collisions. $T=10$, $\alpha=10$ ($\tau_f \approx 2$), and
$\tau =0.1$. The number of particles is  $N=500$ and the final time
$500$. b) Same as before but with almost elastic collisions and
$\tau_c <<\tau_f$. Here $r=0.99$, $\tau_c=0.01$, $\tau_f \approx 2$. 
c) Same as before but with inelastic collisions, $r=0.6$.
d)  Also the case
of  inelastic collisions, but now $\tau_f <<\tau_c$.
The number of particles is $N=200$ and $T=0.2$, $\alpha=1$ ($\tau_f \approx 0.08$),
$\tau=0.2$, $\tau_c=0.1$, and $r=0.1$. Final time also $500$ units.
}
\label{fig:noinertia}
\end{figure} 

\begin{figure}
\begin{center}
\includegraphics[clip=true,width=\columnwidth, keepaspectratio]{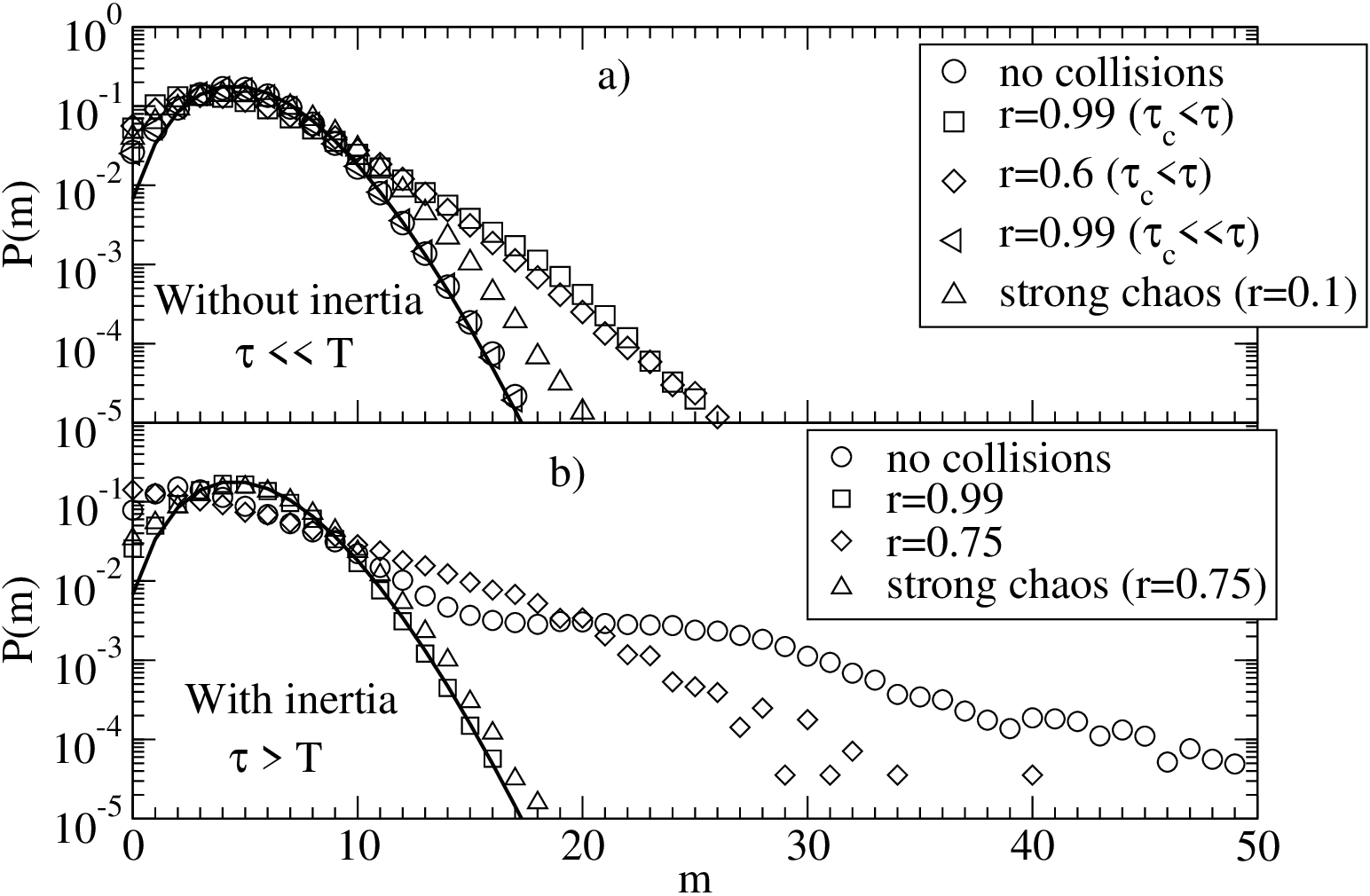}
\end{center}
\caption{Distribution of clusters for the same cases shown in figure 1 (above) and 3 (below).
Solid line is the analytical Poisson distribution.}
\label{fig:pdm}
\end{figure}

First we consider the case of  irrelevant inertia,  $\tau << T$. Obviously, in
the absence of collisions, particles follow strictly the fluid
and no aggregation appears (see Fig.~\ref{fig:noinertia}a)). In
fact, $\tau$ is the relaxation time of the particles to the driving
flow; without collisions the particles are continuously flowing through
the entire spatial domain since no KAM tori are present in the
flow. However, everything changes when the collisions
are taken into account. Thus, if $\tau_c < \tau$, the last term in
Eq.~(\ref{adim}) is the most relevant one.  If we have also that $\tau_c <
\tau_f$ an aggregation of the particles is always observed for any
value of $r$ (even, and most noticeably, at $r=1$). This can be
understood from Eq.~(\ref{adim}), since now the velocity of the
particles can be written as:
\begin{equation}
\hat \vp \approx \hat \vf +\frac{\tau}{\tau_c}\bf F_{col}.
\label{velinertia}
\end{equation}
Thus, the effect of collisions is similar to that of inertia when considered
alone: {\it the
velocity of the particles is modified respect to that of the fluid
parcels, then it is no longer compressible 
($\bf \nabla \cdot \hat \vp \neq 0$), and there are regions where
particles aggregate}. Figs.~\ref{fig:noinertia}b) and 
\ref{fig:noinertia}c)  show a pattern of the
distribution of particles for $r=0.99$ and $r=0.6$, respectively. It is also very
important to note here a specific behavior of elastic or quasi-elastic
collisions ($r \approx 1$). The effect of multiple elastic collisions
is equivalent to macroscopic diffusion, which, if large enough, may
induce a dispersing mechanism that prevents clustering.  Summing up,
one can say that elastic collisions induce a clustering inertia-like
effect, that disappears when diffusion is strong
enough. Quantitatively this is controlled by the adimensional
parameter $\tau/\tau_c$ since the macroscopic diffusivity $D \propto
1/\tau_c$. Increasing $\tau/\tau_c$ clustering like that of
Fig.~\ref{fig:noinertia}b) disappears and a homogeneous distribution
is attained (see discussion below on the $P(m)$ function and
Fig.~\ref{fig:pdm}a)).

On the contrary, when $\tau_f < \tau_c$ (and also $\tau_c << \tau$)
the strong chaoticity of the flow dominates and avoids the formation
of any clustering in the system, see  
 Fig.~\ref{fig:noinertia}d).
Finally, it is obvious from Eq.(\ref{adim}) that the limit $\tau
<<\tau_c$ is equivalent to the absence of collisions, thus
 no clusterization appears.

A more quantitative analysis of clustering follows.  This is performed
via a particle-in-cell histogram, i.e., after dividing the system in
$M$ small boxes (we use $N/M=5$) an histogram is made with the number
of boxes containing $m$ particles, denoted the corresponding function
as $P(m)$. For an homogeneous system of particles $P(m)$ is a Poisson
distribution $exp(-\lambda)\lambda^m/m!$, where $\lambda=N/M$. As the
clustering is stronger developed, the deviations from a Poissonian are
more evident. Fig.~\ref{fig:pdm}a) calculates $P(m)$ for the patterns
shown in Fig.~\ref{fig:noinertia}, and one more case.  
We observe the appearance of clustering for $r<1$, and even for $r \approx 1$. Also,
it is shown how for the elastic case the Poisson distribution is approached when
$\tau/\tau_c <<1$.

\begin{figure}[htb]
\begin{center}
\includegraphics[clip=true,width=\columnwidth, keepaspectratio]{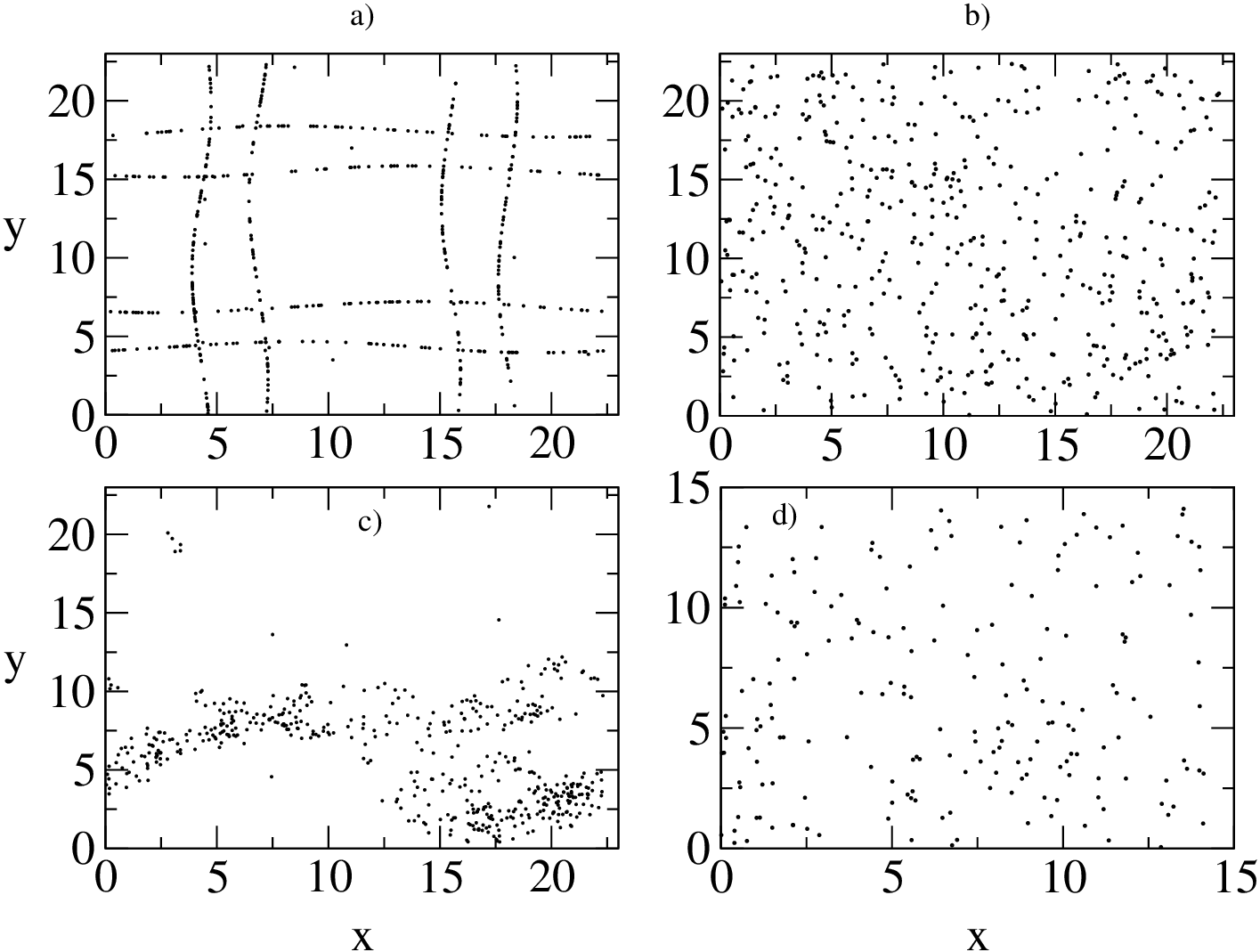}
\end{center}
\caption{Spatial distribution of particles in the steady state, when
the inertia is relevant, $\tau >T$. 
a) Corresponds to the case without collisions.  Here
$T=1$, $\alpha=0.5$,  
and $\tau=10$. The number of particles is $N=500$ and the final time
$500$. b) Same as before but now with almost elastic collisions, with $\tau_c <<\tau_f$.
Here $r=0.99$, $\tau_c=0.1$, $\tau_f \approx 1$. c) Same as before
but with inelastic collisions, $r=0.75$ (here the final time is $175$ as the
program breaks because of collapse earlier. d) Also the case
of  inelastic collisions, but now $\tau_f <<\tau_c$.
The number of particles is $N=200$ and $T=0.2$, $\alpha=1$ ($\tau_f \approx 0.08$),
$\tau=2$, $\tau_c=0.2$, and $r=0.75$. Final time also $500$ units.
}
\label{fig:inertia}
\end{figure} 


It is very relevant to know the effect of collisions in the context of
transport of inertial particles in chaotic flows.  Therefore, we next
consider the general case where the inertial term is relevant, $\tau >
T$ (but otherwise $\tau$ small enough such that the l.h.s is not
negligible with respect to first term in the r.h.s). It is
well-known~\cite{gen:tur} that, in the absence of collisions, heavy
particles accumulate in regions of low strain and high vorticity, the
so-called preferential concentration phenomenon.  In our case, the
clustering steady pattern is shown in Fig.~\ref{fig:inertia}a). After
including collisions one has: if $\tau << \tau_c$ no effect is
observed as the last term in the r.h.s of Eq. (\ref{adim}) can be
neglected. A different behavior is observed when $\tau_c << \tau$.
Regardless of the value of $\tau_f$, in the limit of elastic
collisions ($r \approx 1$) no aggregation of the particles is
observed, Fig.~\ref{fig:inertia}b).  As explained before, when
$\tau/\tau_c$ is large, the continuous elastic colliding processes of
the particles induce a diffusive effect that prevents clustering.  On
the contrary, as we go into the inelastic regime, $r<1$, the {\it
chaotic} time scale, $\tau_f$, now plays a very important role. Thus,
when $\tau_c <
\tau_f$ particles aggregate although in a very different pattern to
that obtained in the absence of collisions (Fig.~\ref{fig:inertia}c)).
In particular, at variance with clustering in the absence of
collisions (see Fig.~\ref{fig:inertia}a)), clusters are created and
destroyed continuously, and moving with the flow.  In this case, the
particles aggregate thanks to the inelastic collisions and because the
flow is not chaotic enough to disperse completely the clusters. The
opposite is observed, Fig.~\ref{fig:inertia}d), when $\tau_f <
\tau_c$; the clusters do not resist the dispersion of the flow and the
density of particles is homogeneous.
The corresponding $P(m)$ function is plotted in
Fig.~\ref{fig:pdm}b). 

For completeness, we briefly mention the limit $\tau >> T$.  This is
completely equivalent to a freely cooling granular gas since the
external flow is irrelevant. Thus, clustering is usually observed for
inelastic collisions~\cite{goldhirsch}.

In this letter we have studied numerically the stationary spatial
structure of a granular material advected by a chaotic flow. We have
seen that the relative importance of collisions, inertia and
chaoticity of the flow gives rise to transitions from homogeneous to
inhomogeneous density of grains. We think that our simple model is
relevant for real physical phenomena that involve the transport of a
large amount of heavy particles, like for example those appearing in many
geophysical situations.

\begin{acknowledgments}
We acknowledge fruitful discussions with Andrea Baldassarri, Fabio
Cecconi, Umberto Marini Bettolo Marconi, and Angelo
Vulpiani, and Emilio Hern\'andez-Garc\'\i a
for a critical reading of the manuscript.  C.L. acknowledges support from the Spanish
MECD. A.P. acknowledges  support from the INFM Center for
Statistical Mechanics and Complexity (SMC).
\end{acknowledgments}





\begin{thebibliography}{99}
\bibitem{note}
It should be noted that equations~\eqref{modelo} represent a
dissipative dynamical system, i.e. even in the case of non-interacting
(or equivalently elastic) particles an attractor is always reached.
\end{thebibliography}
\end{document}